\documentclass[10pt,journal,compsoc]{IEEEtran}
\ifCLASSINFOpdf
\else
\fi
\usepackage[space,nocompress]{cite}
\usepackage{graphicx,caption,amssymb,amsmath,array}
\usepackage{color,float}
\usepackage{subfigure,epsfig}
\usepackage{ragged2e}
\usepackage{enumitem}
\usepackage{footnote}

\usepackage{algorithm}
\usepackage{algorithmic}
\usepackage{multirow}
\usepackage{float}
\usepackage{url}
\usepackage{enumitem}
\usepackage{array}
\usepackage{booktabs}
\usepackage{bm}
\usepackage{color}

\setlist[itemize]{leftmargin=*}
\setlist[enumerate]{leftmargin=*}
\begin{document}
%
\title{Micro-video recommendation model based on graph neural network and attention mechanism}

\author{Chan Ching Ting, Mathew Bowles, Ibrahim Idewu
}

\markboth{Journal of \LaTeX\ Class Files,~Vol.~14, No.~8, August~2015}%
{Shell \MakeLowercase{\textit{et al.}}: Bare Demo of IEEEtran.cls for IEEE Transactions on Magnetics Journals}
%



\IEEEtitleabstractindextext{%
\justifying  
\begin{abstract}
With the rapid development of Internet technology and the comprehensive popularity of Internet applications, online activities have gradually become an indispensable part of people's daily life. The original recommendation learning algorithm is mainly based on user-microvideo interaction for learning, modeling the user-micro-video connection relationship, which is difficult to capture the more complex relationships between nodes. To address the above problems, we propose a personalized recommendation model based on graph neural network, which utilizes the feature that graph neural network can tap deep information of graph data more effectively, and transforms the input user rating information and item side information into graph structure, for effective feature extraction, based on the importance sampling strategy. The importance-based sampling strategy measures the importance of neighbor nodes to the central node by calculating the relationship tightness between the neighbor nodes and the central node, and selects the neighbor nodes for recommendation tasks based on the importance level, which can be more targeted to select the sampling neighbors with more influence on the target micro-video nodes. The pooling aggregation strategy, on the other hand, trains the aggregation weights by inputting the neighborhood node features into the fully connected layer before aggregating the neighborhood features, and then introduces the pooling layer for feature aggregation, and finally aggregates the obtained neighborhood aggregation features with the target node itself, which directly introduces a symmetric trainable function to fuse the neighborhood weight training into the model to better capture the different neighborhood nodes' differential features in a learnable manner to allow for a more accurate representation of the current node features. The model uses graph neural networks to efficiently model user preference information through information transfer and information aggregation on nodes, while introducing attention networks to finally obtain rating predictions.
\end{abstract}

}

\maketitle

\IEEEdisplaynontitleabstractindextext

%
\IEEEpeerreviewmaketitle

\section{Introduction}
 With the gradual development of Internet technology, information resources are growing at a high speed and the problem of "information overload" has emerged. It is difficult for users to get the information they need directly from the huge amount of information. Therefore, recommendation algorithms are usually used to solve the problem of "information overload", and the research of various recommendation methods~\cite{sedhain2015autorec, cheng2016wide, salah2018probabilistic} has become one of the main focuses in the field of computer science. 

The traditional recommendation system based on collaborative filtering~\cite{he2016fast} has been the main direction of recommendation algorithm research because it can take the advantage of big data to find users with the most similar interests and recommend the micro-videos they are interested in. However, there are two problems that are difficult to solve in collaborative filtering, namely, the cold start problem caused by too little initial data of new users and the sparse interaction data problem caused by too many micro-videos in total and too few micro-videos interacting with users. To solve these two problems, researchers have tried to introduce various kinds of auxiliary information to complement the data, such as point-of-interest letters~\cite{cao2010mining}, comment information~\cite{zhang2013generating}, social networks ~\cite{jamali2010matrix}, and contextual information~\cite{adomavicius2011context}, which have been useful in specific scenarios, and researchers have started to explore how to design better models to fully utilize the auxiliary information and explore the applicability of The researchers started to explore how to design better models to fully utilize the auxiliary information and explore more general auxiliary information with wider applicability~\cite{wang2019neural}. Recently proposed approaches prefer to design end-to-end models to simultaneously learn the potential feature attributes and structural relationships of entities in the knowledge graph and use them for recommendation, which are implemented mainly by graph neural network~\cite{scarselli2008graph} (GNN) approaches to model the knowledge graph.

Compared with traditional graph learning, graph neural networks can not only learn the topology of graph networks but also aggregate feature information of neighbors, thus enabling effective learning of various structures in graph networks and playing a key role in subsequent recommendations~\cite{defferrard2016convolutional}. The original recommendation learning algorithm~\cite{mnih2007probabilistic} is mainly based on user-micro-video interaction for learning and modeling the user-micro-video connectivity, which makes it difficult to capture more complex relationships between nodes. While traditional graph learning generally works for graph topology and also takes less into account various feature information among nodes or nodes, compared with the original recommendation methods, graph neural network recommendation methods can alleviate the problems of cold start~\cite{lee2016enhancing,wei2021contrastive} and data sparsity~\cite{zhou2011functional} in traditional collaborative filtering methods. It can not only learn the topology of the graph network but also aggregate the various connections of neighboring nodes. Thus it can learn the information in the graph network more effectively and play a key role in the subsequent recommendation work.

Compared with traditional methods, graph neural networks can improve the accuracy of recommendation results. The application scenarios of recommendation systems often contain huge amounts of data, and traditional collaborative filtering recommendations often fail to make recommendations for users because they have too little interaction data, or a large number of users have high similarity to them due to a lot of interaction data of some users, which leads to less personalized recommendation results. For example, if user A likes movies 1, 2 and 3, and user B likes movie 1, the recommendation system will consider that user A and user B are similar and obtain the relationship that both 2 and 1 are the works of director c from the knowledge graph, and then give priority to The graph neural network can improve the diversity ~\cite{smith2019towards} and novelty~\cite{zhou2010solving} of recommendation results.

Traditional recommendation algorithms tend to have a large number of repetitive recommendations as well as high similarity recommendations, which result in a lack of novelty in the recommended results due to the single index for calculating similarity~\cite{Yu}. The inferable implicit information contained in the knowledge graph enhances the ability of the knowledge graph-based recommendation system to explore the potential interests of users and provide them with richer recommendation results. For example, if user A likes movies 4 and 5, the recommendation system can recommend 6 for user A based on the relationship that both 4 and 6 are the works of director d, and recommend 7 for user A based on the relationship that both 7 and 1 are starred by e. The graph neural network has good interpretability. Since the nodes of the knowledge graph have realistic meaning, they can be linked with the user's history to provide a basis for the user's recommendation. Graph neural networks have good scalability and wider applicability~\cite{NMCL}. The knowledge graph-based recommendation algorithm relies on the objective knowledge contained in the knowledge graph as well as the implicit derivable knowledge; in other words, the more knowledge the knowledge graph contains, the more powerful the knowledge graph-based recommendation algorithm is. Monti et al.~\cite{monti2017geometric} used graph neural networks to extract network representations of users and micro-videos, and then combined with recurrent neural networks for the message passing process. Berg et al.~\cite{berg2017graph} proposed a graph auto-coding framework to generate latent features of users and micro-videos by passing message aggregation on the user-micro-video graph.Hartford et al.~\cite{hartford2018deep} considered the problem of predicting relationships between two and more different sets of objects and introduced a weight binding scheme to the deep model.

The original recommendation learning algorithm is mainly based on user-micro-video interaction for learning, modeling the user-micro-video connection relationship, which is difficult to capture the more complex relationships between nodes. And traditional graph learning generally works for graph topology and also takes less account of various feature information among nodes and nodes~\cite{GRCN}. To address the above problems, this paper proposes a graph neural network-based model. We propose a personalized recommendation model based on graph neural network, which utilizes the feature that graph neural network can tap deep information of graph data more effectively, and transforms the input user rating information and micro-video side information into graph structure, ( including user-micro-video graph, micro-video-information graph, etc.) for effective feature extraction, based on the importance sampling strategy. The importance-based sampling strategy measures the importance of neighbor nodes to the central node by calculating the relationship tightness between the neighbor nodes and the central node, and selects the neighbor nodes for recommendation tasks based on the importance level, which can be more targeted to select the sampling neighbors with more influence on the target micro-video nodes. The pooling aggregation strategy, on the other hand, trains the aggregation weights by inputting the neighborhood node features into the fully connected layer before aggregating the neighborhood features, and then introduces the pooling layer for feature aggregation, and finally aggregates the obtained neighborhood aggregation features with the target node itself, which directly introduces a symmetric trainable function to fuse the neighborhood weight training into the model to better capture the different neighborhood nodes' differential features in a learnable manner to allow for a more accurate representation of the current node features. The model uses graph neural networks to efficiently model user preference information through information transfer and information aggregation on nodes, while introducing attention networks to finally obtain rating predictions.

\begin{figure}
	\centering
	  \includegraphics[width=0.5\textwidth]{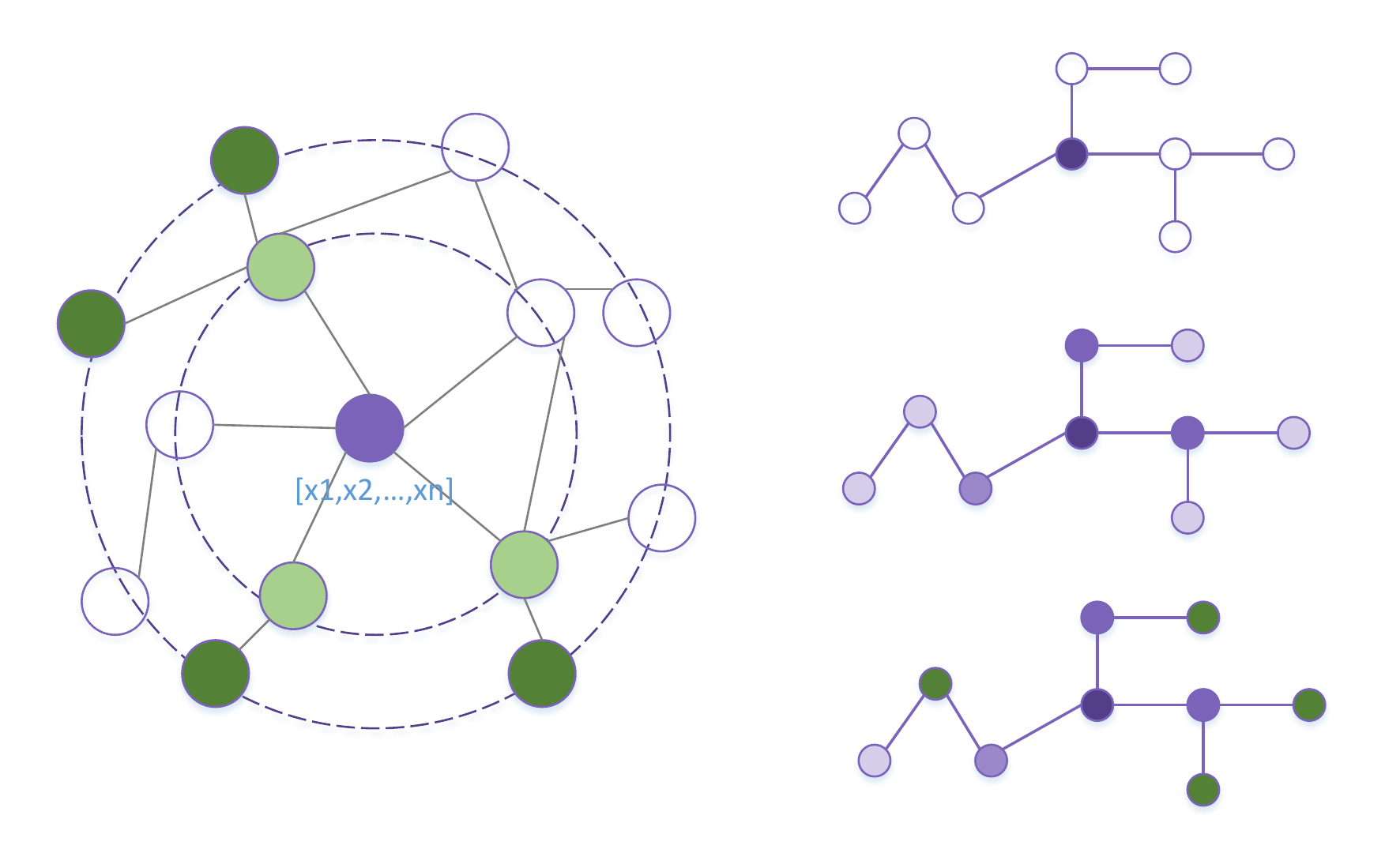}
	  \vspace{-1mm}
	  \caption{Schematic diagram of multi-layer convolutional graph neural network.}
	\label{fig_1}	  
	\vspace{-5mm}
\end{figure}

\section{Related Work}
Graph neural networks were first proposed by Gori et al. in 2005~\cite{scarselli2005graph} and then further elucidated by Scarselli et al. in 2009~\cite{scarselli2008graph}. In early studies based on graph data, for the target node representation, researchers conducted learning by iteratively propagating neighbor information through recurrent neural networks. The computational effort in this process is very large, and many researchers hope to alleviate this problem, and subsequently this has become a research hotspot in recent years. The relationship between graph neural networks and network embeddings is close, and the relationship between them is briefly described here. Network embedding is the projection of nodes in a graph into a low-dimensional vector space while preserving the network topology and node information. thus producing a low-dimensional vector representation of the nodes, which is then used for graph data analysis (e.g., classification, recommendation, etc.). The general network embedding algorithms can be broadly classified into: matrix decomposition~\cite{koren2009matrix,xue2017deep,mnih2007probabilistic}, random wandering ~\cite{perozzi2014deepwalk} and deep learning methods~\cite{wang2016structural,hamilton2017inductive}. Among them, deep learning methods for network embedding also belong to graph neural networks, including algorithms based on graph self-encoders (e.g., SDNE~\cite{wang2016structural}) and graph convolutional neural networks with unsupervised learning (GraphSage~\cite{hamilton2017inductive}),and LightGCN~\cite{he2020lightgcn}.

GC-MC~\cite{berg2017graph} is a graph-based auto-encoder framework for matrix completion. The auto-encoder produces latent features of user and item nodes through a form of message passing on the bipartite interaction graph. These latent user and item representations are used to reconstruct the rating links through a bilinear decoder. The benefit of formulating matrix completion as a link prediction task on a bipartite graph becomes especially apparent when recommender graphs are accompanied with structured external information such as social networks. Combining such external information with interaction data can alleviate performance bottlenecks related to the cold start problem.

Spectral graph theory~\cite{shuman2013emerging} studies connections between combinatorial properties of a graph and the eigenvalues of matrices associated to the graph. L. Zheng et al. ~\cite{zheng2018spectral} proposed a spectral graph theory based method to leverage the broad information existing in the spectral domain. Specifically, to conquer the difficulties of directly learning from the spectral domain for recommendations, they first presented a new spectral convolution operation, which is approximated by a polynomial to dynamically amplify or attenuate each frequency domain. Then, they introduced a deep recommendation model, named Spectral Collaborative Filtering (SpectralCF), built by multiple proposed spectral convolution layers. SpectralCF directly performs collaborative filtering in the spectral domain.
 
  J. Sun et al.~\cite{sun2020neighbor} propose a novel graph convolutional neural network-based recommender system framework, Neighbor Interaction Aware Graph Convolutional Neural Networks, NIA-GCN. In NIA-GCN, they proposed a Pairwise Neighborhood Aggregation (PNA) layer to capture relationships between pairs of neighbors at each GCN layer. PNA applies element-wise multiplication between every two neighbor embeddings to learn user-user and item-item relationships, which are important signals in recommendation. Instead of recursively updating neighborhood embeddings, they introduced parallel graph convolution networks (Parallel-GCNs), which can independently aggregate information from node entities at different depths with respect to the central node. Such an approach is a much better match to the heterogeneous nature of the user-item bipartite graph. They proposed a novel Cross-Depth Ensemble (CDE) layer to capture the user-user, item-item and user-item relationships in neighborhoods in the graph. This layer allows predictions to take into account more complicated relationships across different depths in the graph.
  
  S. Ji et al.~\cite{ji2020dual} proposed a dual channel hypergraph collaborative filtering (DHCF). It is a dual channel learning strategy, which holistically leverages the divide-and-conquer strategy, is introduced to learn the representation of users and items so that these two types of data can be elegantly interconnected while still maintaining their specific properties. The hypergraph structure is employed for modeling users and items with explicit hybrid high-order correlations. The jump hypergraph convolution (JHConv) method is proposed to support the explicit and efficient embedding propagation of high-order correlations. Wang et al.~\cite{GRCN} constructed the user-item graph, whose edges corresponded to implicit feedback. With their proposed neural graph collaborative filtering (NGCF) method, the collaborative signal conveyed by the edges and high-order connectivity explicit modelled and injected into each user and item embedding. More recently, the GCN-based model has been introduced into multimedia recommendation in implicit feedback settings. Wei et al.~\cite{MMGCN} constructed the modal-specific bipartite graph with implicit data to model the user preference in multiple modalities. They developed a multimedia recommendation framework, dubbed multimodal graph convolutional network (MMGCN), which represented the user preference in each modality with her/his directly and indirectly connected neighbors.

\section{Methodology}
\subsection{Embedding Layer}

\begin{figure*}
	\centering
	  \includegraphics[width=1\textwidth]{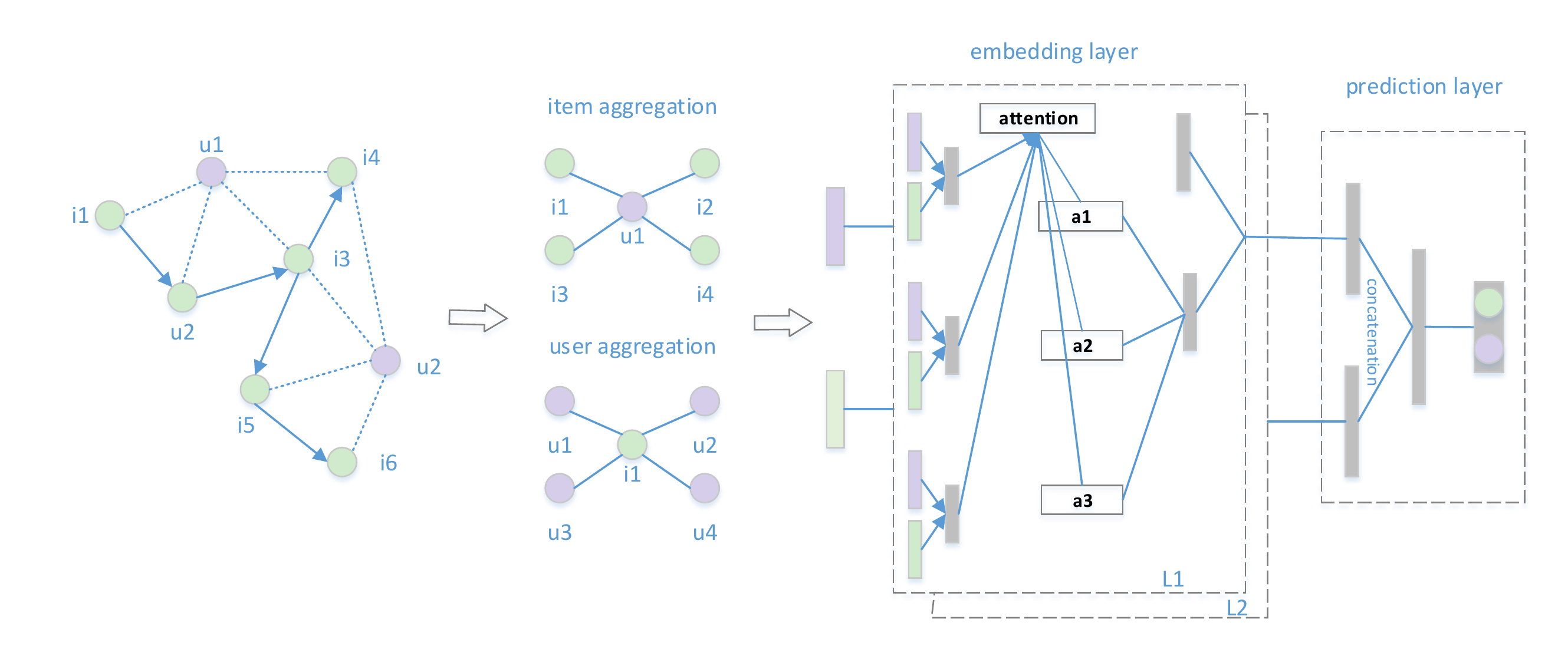}
	  \vspace{-1mm}
	  \caption{Schematic diagram of our neural network recommendation model.}
	\label{fig_1}	  
	\vspace{-5mm}
\end{figure*}

We use an embedding vector $e_{ua} \in \mathbb{R}^d$ to describe a user $u_a$, where $d$ denotes the size of the embedding. This can be viewed as constructing the parameter matrix as an embedding lookup table, as the following equation shown:
\begin{equation}
E=[e_{ul}, \dots, e_{um}, e_{il}, \dots, e_{im}, e_{1}, \dots, e_{n}],
\end{equation}
where $e_{ul}, \dots, e_{um}, e_{il}$ represents user embedding, $e_{il}, \dots, e_{im}, e_{1}$ represents micro-video embedding. $e_{il}, \dots, e_{im}, e_{1}$ represents rates. In the GCFA framework, we jointly consider user embedding, micro-video embedding and rate embedding, feed the end-user embedding and micro-video embedding into a multilayer perceptron, and then output the predicted of the scores.

Create the user embedding matrix $U$ and the project embedding matrix $V$. Initialize using the common initializer Xavier, which keeps the gradient size of each layer approximately the same. Then input the users in turn, and the user adjacency matrix $N$ to aggregate the user neighbor feature adjacencies as follows:
\begin{equation}
u^n=\sum_{n\in N(u)} g(r_{u,n})n,
\end{equation}
where $N(u)$ is the set of users' neighbors and $g(r_{u,n})$ is the normalization of the relationship weights between users and their neighbors, as shown in follows:
\begin{equation}
g(r_{u,n})=\frac{exp(r_{u,n})}{\sum_{n\in N(u)} exp(r_{u,n})},
\end{equation}
Then the aggregated neighborhood features are nonlinearly aggregated with the user's own features, as shown below:
\begin{equation}
agg_u=\delta(w\cdot (u+u^n)+b),
\end{equation}
where $\delta$ is the nonlinear activation function. We use Relu as the activation function, $w$ is the transformation weight, and $b$ is the bias term.

\subsection{Propagation Layer}

For a connected user-micro-video pair, we define the message propagate from $i_s$ to $u_a$ as:
\begin{equation}
h_a=\theta(w\cdot Aggre_{i}(\{x_{sa}, \forall{s}\in C(a)\}+b)),
\end{equation}
where $C(a)$ denotes the set of micro-videos that user $u_a$ has interaction with,  $x_{sa}$ denotes the interaction between user $u_a$ and micro-video $i_s$. Based on the embedding representation, $Aggre_{i}$ is the micro-video aggregation function. In addition $\theta$ denotes the nonlinear activation function and $w$, $b$ are the weights and biases of the neural network, respectively.

For the interaction between user $u_a$ with rating level $r$ and micro-video $i_s$, a two-layer neural network is used to combine the micro-video embedding $e_i$ with the viewpoint embedding $e_r$, thus modeling the viewpoint-based representation $x_{as}$ of the interaction. This can be expressed as g with the tandem of micro-video embedding and viewpoint embedding as input and the output as The fusion network for the interaction viewpoint-based representation. $x_{as}$ is represented as follows:
\begin{equation}
x_{as}=g_v ([e_i \oplus e_r]),
\end{equation}
where $\oplus$ denotes the concatenation of two vectors.

The traditional aggregation operation $Aggre_{i}$ is based on the mean operator, which takes the mean value of the elements of the vector in $\{x_{as}, \forall{s}\in C(a)\}$. The mean value-based aggregation function is as follows:
\begin{equation}
h_a=\theta(w\cdot \sum_{s\in C(a)} \beta_a x_{as} +b),
\end{equation}
In the mean-based aggregation, $\beta_a$ is fixed to $\frac{1}{C(a)}$ for all micro-vide0s. The approach assumes that all interactions contribute equally to understanding the user. In fact, not every interaction behavior has the same impact on user based viewpoint embedding, so we allow various interactions to contribute differently to the user's latent factor. Assign unique weights to each $(u_a, i_s)$ pair:
\begin{equation}
h_a=\theta(w\cdot \sum_{s\in C(a)} \beta_{as} x_{as} +b),
\end{equation}

When understanding user preferences from historical interactions $C(a)$, $ \beta_{as}$ denotes the attentional weight of the contribution of the interaction with $i_s$ to the potential factor of user $u_a$ when understanding the potential factor of user $u_a$. In particular, we parameterize micro-video attention $ \beta_{as}$ using an attention network consisting of two layers of networks. The inputs to the attention network are the interaction-based viewpoint representation $x_{as}$ and the embedding $e_{ua}$ of the target user $u_a$. The attention network is defined as follows:
\begin{equation}
\beta_{as}^*=w_2^T \cdot \theta ([x_{as} \oplus e_{ua}]+b_1)+b_2,
\end{equation}

The above attention scores are normalized using the Softmax function to obtain the target attention weights such that the contribution to the user potential factor of user $u_a$ is:
\begin{equation}
\beta_{as}=\frac{exp(\beta_{as}^*)}{\sum_{s\in C(a)} exp(\beta_{as}^*)},
\end{equation}
Substitute equation(10) into equation(8) to calculate $h_a$.

Improving the representation of $u_a$ by aggregating information from user's neighbors. Concatenate $h_a$ and $e_{ua}$. The results are fed into a two-layer neural network, and the first layer embedding is obtained through the attention mechanism and the embedding propagation layer to optimize the optimize the initial embedding.
\begin{equation}
e_{ua}^{(l)}=LeakyReLU(m_{u\gets u}+\sum_{s\in C(a)} m_{u\gets i}),
\end{equation}
The activation function of LeakyReLU allows messages to encode small signals, both positive and negative. Similarly, we can obtain a representation of the micro-video $e_{ib}^{(l)}$ by propagating information about its associated users $i_b$.
\begin{equation}
e_{ib}^{(l)}=LeakyReLU(m_{i\gets i}+\sum_{i\in B(b)} m_{i\gets u}),
\end{equation}
In summary, the advantage of embedding propagation layers is the explicit use of single-layer connection information to associate users and micro-video representations.

\subsection{Loss Function}
The loss function of the model is shown below:
\begin{equation}
\begin{split}
& L=\sum_{(u,v)\in M} -log\theta(u_a^T i_s)+ log(1-\theta(u_a^T i_s)) \\
& +\frac{\lambda}{2} (\|U\|^2_2 + \|V\|^2_2 +\|W\|^2_2,    
\end{split}
\end{equation}
The former term is the cross-entropy loss between the true and predicted scores, and the latter term is a regularization term used to avoid overfitting due to large parameter differences.
\section{EXPERIMENTS}
\subsection{Dataset}
 \begin{figure*}
    \centering
    \subfigure[AUC on Movielens]{
      \includegraphics[width=0.45\textwidth]{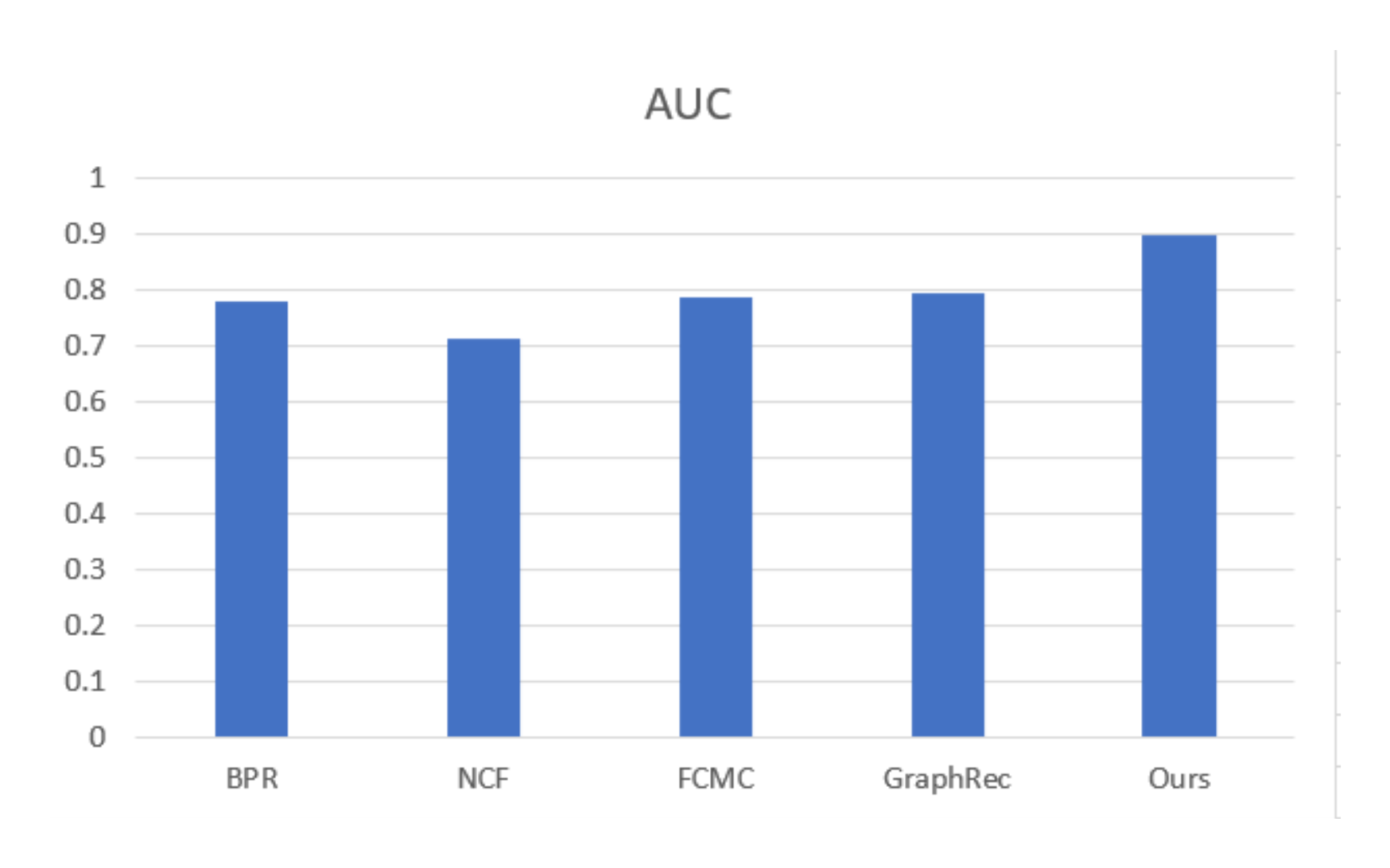}
      \label{fig_visualize_1_1}
    }
    \subfigure[AUC on Amazon]{
      \includegraphics[width=0.45\textwidth]{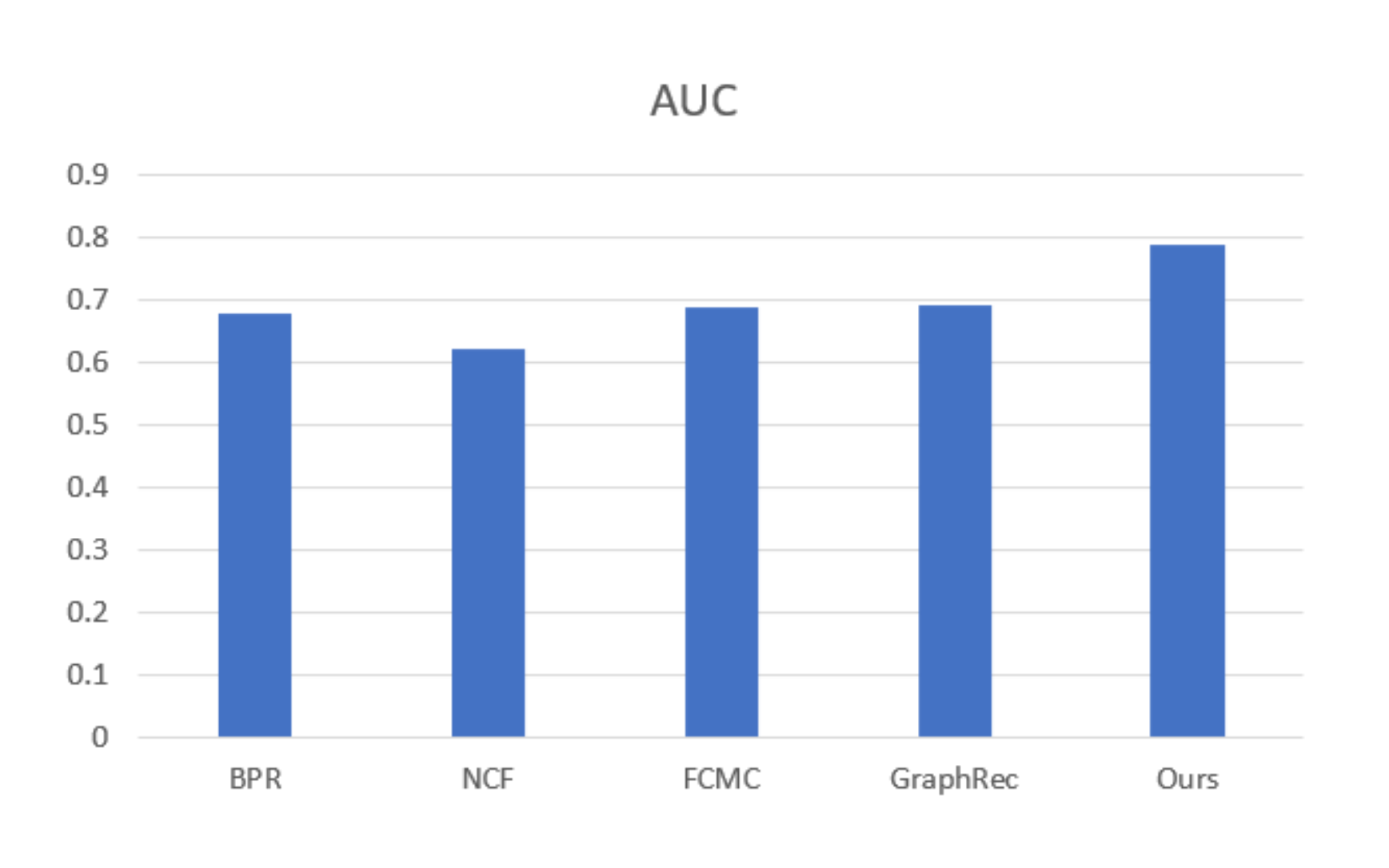}
      \label{fig_visualize_2_3}
    }

    \vspace{-10pt}
    \caption{AUC Comparison Histogram.}
    \label{fig_3}
    \vspace{-5pt}
 \end{figure*} 
 
\begin{table}
  \centering
  \caption{ecommendation Accuracy Performance Comparison.}
  \label{table_3}
  \setlength{\tabcolsep}{5.0mm}
  \begin{tabular}{|c|c|c|c|}
    \hline
    \textbf{Model}&\textbf{AUC}&\textbf{NDCG@1}&\textbf{NDCG@2}\\
    \hline
    \textbf{BPR}&$0.7345$&$0.7536$&$0.9213$\\
    \hline
    \textbf{NCF}&$0.7436$&$0.7569$&$0.9311$\\
    \hline
    \textbf{FCMC}&$0.7497$&$0.7641$&$0.9422$\\
    \hline
    \textbf{GraphRec}&$0.7536$&$0.7753$&$0.9495$\\
    \hline
    \textbf{Ours}&$0.7721$&$0.7938$&$0.9567$\\
    \hline
    
  \end{tabular}
  \vspace{-2mm}
\end{table}

The datasets used in this experiment are MovieLens-1M and Amazon.

MovieLens-1M~\cite{harper2015movielens}: This movie rating dataset has been widely used to evaluate collaborative filtering algorithms. We used the version containing 1,000,209 ratings from 6,040 users for 3,900 movies. While it is a dataset with explicit feedbacks, we follow the convention~\cite{he2017neural} that transforms it into implicit data, where each entry is marked as 0 or 1 indicating whether the user has rated the micro-video. After transforming, we retain a dataset of $1.0\%$ density.

Amazon Instant Video~\cite{mcauley2015image}: The dataset consists of 426,922 users, 23,965 videos and 583,933 ratings from Amazon.com. Similarly, we transformed it into implicit data and removed users with less than 5 interactions. As a result, a dataset of $0.12\%$ density is obtained. 

\subsection{Baselines}

BPR~\cite{rendle2012bpr}: We use Bayesian Personalized Ranking based Matrix Factorization. BPR introduces a pair-wise loss into the Matrix Factorization to be optimized for ranking.

NCF~\cite{he2017neural}: Neural Collaborative Filtering fuses matrix factorization and Multi-Layer Perceptron (MLP) to learn from user-item interactions. The MLP endows NCF with the ability of modelling non-linearities between users and items. 

GCMC~\cite{berg2017graph}: Graph Convolutional Matrix Completion utilizes a graph auto-encoder to learn the connectivity information of a bipartite interaction graph for latent factors of users and items.

GraphRec~\cite{fan2019graph}: this model uses GNNs to integrate node information and topology for consistent modeling of user-project graphs and user-user social graphs.

\subsection{Evaluation Metrics}

The most commonly used metrics for video recommendation and even general-purpose recommendation systems are ranking metrics, of which the most representative metrics are the area under the feature curve (AUC) and the normalized discounted cumulative gain (NDCG)~\cite{schutze2008introduction}. The AUC measures the ability of the model to discriminate between all positive and negative samples relative to each other, and the NDCG measures the distance of the model ranking results from the ideal ranking results.

\subsection{Result Analysis}
We first compared the overall recommendation results We compare the overall recommendation results, as shown in Table 1.
From Table 1, we can see that, compared with the existing models, our proposed model is more effective in terms of AUC, NDCG@1, NDCG@2. The average relative improvement value is about $1.7\%$. For the recommendation system model The improvement value is significant for the recommended system model.

NCF obtains better performance than BPR. Both methods utilize only scoring information. However, NCF is based on a neural network architecture, which shows the power of neural network models in recommender systems. graphRec outperforms GCMC. both methods utilize only graph neural networks. However, GraphRec is based on the embedding propagation architecture, which illustrates the power of the embedding propagation model. The method in this paper outperforms all baseline methods in most cases. In contrast to GraphRec, the model in this paper utilizes attention mechanisms in the user micro-video graph to fuse both interaction behaviors of user micro-videos and user perspectives. This illustrates the importance of simultaneously fusing the interaction behavior of user micro-videos and user viewpoints. In the Book-Crossing dataset, where the data is sparse, the proposed model achieves more obvious advantages compared with all other models, which proves that the improved strategy proposed in this paper is useful in capturing important features and compensating for the sparsity of data. As can be seen from Figure 3, our model achieves a smaller training training loss.
\begin{figure}
	\centering
	  \includegraphics[width=0.5\textwidth]{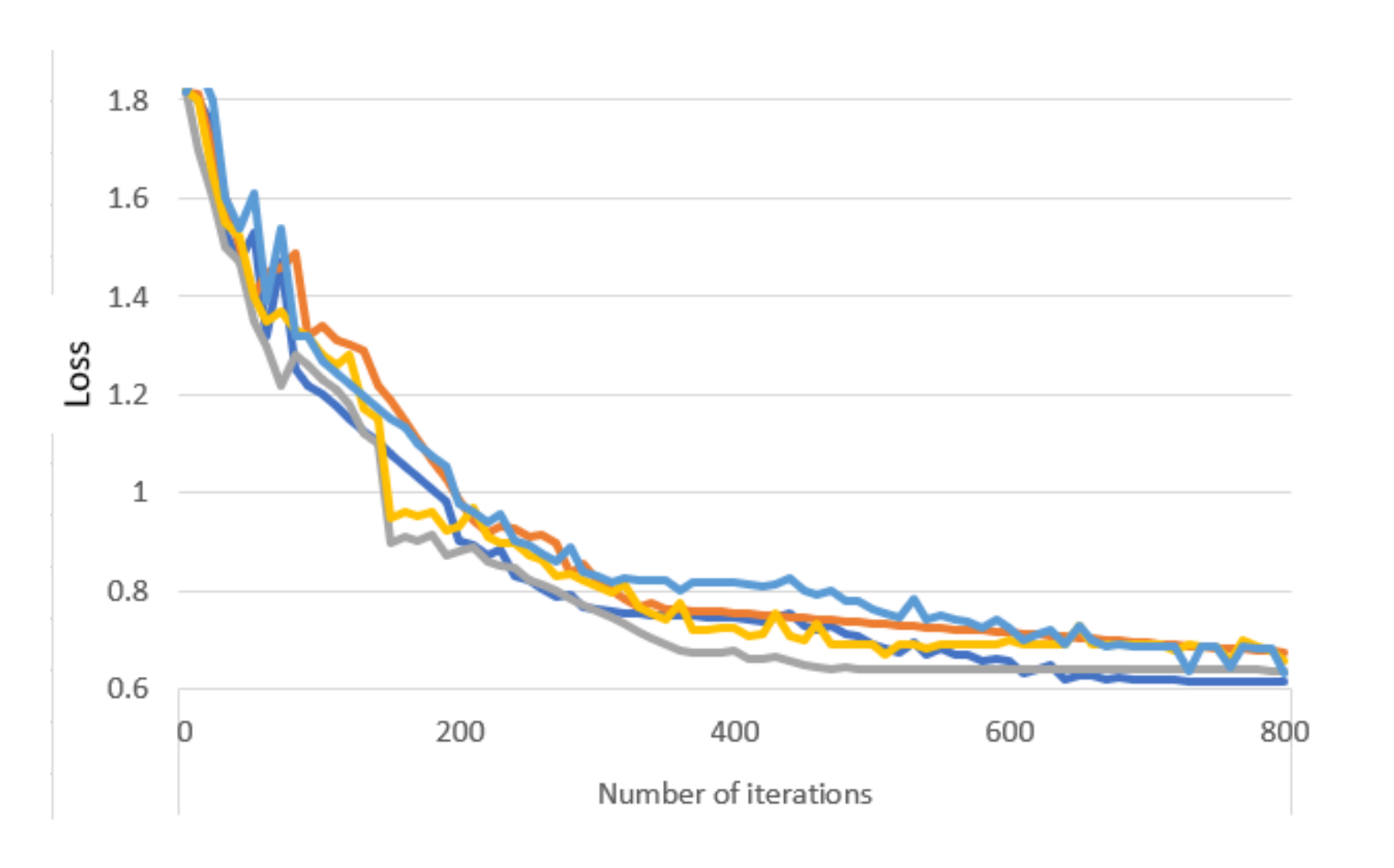}
	  \vspace{-1mm}
	  \caption{Comparison of model loss curves.}
	\label{fig_1}	  
	\vspace{-5mm}
\end{figure}

\section{Conclusion and Future Work}
The original recommendation learning algorithm is mainly based on user-micro-video interaction for learning, modeling the user-micro-video connection relationship, which is difficult to capture the more complex relationships between nodes. And traditional graph learning generally works for graph topology and also takes less account of various feature information among nodes and nodes. To address the above problems, this paper proposes a graph neural network-based model. We propose a personalized recommendation model based on graph neural network, which utilizes the feature that graph neural network can tap deep information of graph data more effectively, and transforms the input user rating information and micro-video side information into graph structure, ( including user-micro-video graph, micro-video-information graph, etc.) for effective feature extraction, based on the importance sampling strategy. The importance-based sampling strategy measures the importance of neighbor nodes to the central node by calculating the relationship tightness between the neighbor nodes and the central node, and selects the neighbor nodes for recommendation tasks based on the importance level, which can be more targeted to select the sampling neighbors with more influence on the target micro-video nodes. The pooling aggregation strategy, on the other hand, trains the aggregation weights by inputting the neighborhood node features into the fully connected layer before aggregating the neighborhood features, and then introduces the pooling layer for feature aggregation, and finally aggregates the obtained neighborhood aggregation features with the target node itself, which directly introduces a symmetric trainable function to fuse the neighborhood weight training into the model to better capture the different neighborhood nodes' differential features in a learnable manner to allow for a more accurate representation of the current node features. The model uses graph neural networks to efficiently model user preference information through information transfer and information aggregation on nodes, while introducing attention networks to finally obtain rating predictions.

\bibliography{BIB/IEEEabrv, reference}

\end{document}